\documentclass[journal]{IEEEtran}
\usepackage{multicol}
\usepackage{cite}
\usepackage[font=small,singlelinecheck=false]{caption}
\usepackage{times}
\usepackage{graphicx}
\usepackage{multirow}
\usepackage[none]{hyphenat}
\usepackage[dvipsnames, x11names,table]{xcolor}
\newcommand{\defcommenter}[2]{%
  \expandafter\newcommand\csname #1\endcsname[1]{%
  {\color{#2}[#1: ##1]}%
  }%
}
\defcommenter{TODO}{blue}
\usepackage{amsmath,amssymb,amsfonts}
\usepackage[T1]{fontenc}
\usepackage{gensymb}
\usepackage{array}
\newcolumntype{L}[1]{>{\raggedright\let\newline\\\arraybackslash\hspace{0pt}}m{#1}}
\newcolumntype{C}[1]{>{\centering\let\newline\\\arraybackslash\hspace{0pt}}m{#1}}
\newcolumntype{R}[1]{>{\raggedleft\let\newline\\\arraybackslash\hspace{0pt}}m{#1}}
\usepackage{float}
\usepackage{subfig}

\usepackage{colortbl}
\begin{document}
%
\title{Fast-Locking and High-Resolution Mixed-Mode DLL with Binary Search and Clock Failure Detection for Wide Frequency Ranges in 3-nm FinFET CMOS}
%
%
%

\author{Nicolás~Wainstein,~\IEEEmembership{Member,~IEEE,}
        Eran~Avitay, and
        Eugene~Avner
\thanks{N. Wainstein was with Intel Corporation, Haifa, Israel. He is currently with the Andrew and Erna Viterbi Faculty of Electrical and Computer Engineering, Technion -- Israel Institute of Technology, Haifa 3200003, Israel.} 
\thanks{E. Avitay, and E. Avner are with Intel Corporation, Haifa, Israel. Corresponding author: e-mail: (nicolasw@technion.ac.il).}
\thanks{Manuscript received XX, 2025; revised XX, 2025.}}

\maketitle

\begin{abstract}
This paper presents a mixed-mode delay-locked loop (MM-DLL) with binary search (BS) locking, designed to cover a broad frequency range from 533 MHz to 4.26 GHz. The BS locking scheme optimizes the locking time, reducing it from a linear to a logarithmic function, completing in B+1 cycles, where 
B represents the digital-to-analog (DAC) resolution controlling the voltage-controlled delay line (VCDL). At the start of the BS process, large step sizes can cause significant bias overshoots, potentially leading to clock failure conditions (\textit{i.e.,} clocks fail to propagate through the VCDL). To address this issue, a toggle detector is introduced to monitor clock activity and adjust the binary search controller. Upon detecting a stalled clock, the controller reverts the DAC code to the previous working code and resumes the BS with a reduced step size. Fabricated in a 3-nm FinFET CMOS process, the proposed MM-DLL achieves a locking time of under 10.5 ns while consuming 5.4~mW from a 0.75~V supply at 4.26~GHz. The measured performance includes a high resolution of 0.73~ps, with a static phase error of 0.73~ps, RMS jitter of 1.2~ps, and peak-to-peak jitter of 4.9~ps. The proposed MM-DLL achieves state-of-the-art power figure of merit (FoM) of 0.82 pJ and DLL locking FoM of 0.01 pJ$\cdot$ns\textsuperscript{2}.

\end{abstract}

\begin{IEEEkeywords}
Delay-locked loop, voltage-controlled delay line, binary search, phase-frequency detector, wireline, deskew, clock generation, clock synthesis, phase error, jitter, locking time.
\end{IEEEkeywords}

%
\IEEEpeerreviewmaketitle

\section{Introduction}
\IEEEPARstart{H}{igh-speed} parallel wireline links are fundamental in any modern system-on-chip (SoC), and at the heart of these interfaces stands the delay-locked loop (DLL). This circuit is specialized in generating one or multiple phases of a reference clock for the deskew systems, and clock and data recovery~\cite{Kim2018,Sidiropoulos1997}. With the increasing data rates of double data rate (DDR) standards and recent advancements in die-to-die (D2D) links for 2.5D and 3D integration~\cite{das2024high,Sharma2022universal,Ardalan2020bunch,Li2024}, ultra-wide-range and low-latency DLLs are required to meet the specifications of new standards while maintaining backward compatibility. Additionally, to achieve day-long battery life in state-of-the-art microprocessors, deep low-power modes are required, making low exit times critical for the overall performance of the SoC. 

In recent years, all-digital DLLs (AD-DLL) have been extensively studied due to their superior scalability with CMOS process, lower power consumption~\cite{Angeli2020,Rehman2020}, and faster locking time~\cite{Park2021,Hossain2014,Kim2021,Wang2010}. However, AD-DLLs struggle to provide the fine resolution required in high-performance wireline links to minimize static phase errors and jitter~\cite{Hsieh2016}. Conversely, analog and mixed-mode DLLs (A-DLLs and MM-DLLs, respectively) can achieve higher operating frequencies and sub-picosecond (ps) resolution by means of highly optimized voltage-controlled delay lines (VCDLs) and phase detectors (PDs)~\cite{Ryu2012,Hossain2014}. This sub-ps time resolution is enabled by fine-grained control of bias or power supply of the VCDL, typically generated by high-resolution digital-to-analog converters (DACs) or charge-pumps (CPs). However, the enhanced temporal resolution in A-DLLs and MM-DLLs comes at the cost of increased locking times since the locking duration is inversely proportional to the time resolution~\cite{Hossain2014}.
Furthermore, to prevent harmonic locking, \textit{i.e.}, locking at an incorrect multiple of the desired delay between the feedback clock (\textit{CLK}\textsubscript{FB}) and the reference clock (\textit{CLK}\textsubscript{REF}), the initial condition of the VCDL bias must ensure that \textit{CLK}\textsubscript{FB} leads \textit{CLK}\textsubscript{REF} at the input of the PD. As a result, the DAC or CP often must be initialized at the minimum code, leading to longer locking times.

A potential solution to long locking times is the use of coarse-fine modes, which adjust the DAC code or capacitive ratios in coarse steps during DLL initialization, until the first polarity change in the phase difference ($\Delta \phi$) between \textit{CLK}\textsubscript{FB} and \textit{CLK}\textsubscript{REF} is detected by the PD~\cite{Park2021}. Once this occurs, the step size is reduced to achieve fine-grained resolution until the locking bias condition is met ($\Delta \phi\approx 0$). While the coarse-fine scheme improves locking times, it can lead to a clock failure condition, where the clocks stop propagating through the VCDL due to large bias overshoots. This condition is irreversible as the PD loses track of the polarity of $\Delta \phi$, necessitating to restart the DLL initialization process with adjusted parameters, such as the coarse step size, or the capacitive load of the delay elements (DEs) within the VCDL. Importantly, in many DLL designs, this issue may create a false locking condition and go undetected until later stages of the initialization flow, potentially affecting subsystems that depend on the output clocks of the DLL.

An extended approach to improve looking time in AD-DLL involves using a time-to-digital converter (TDC) to measure $\Delta \phi$,
allowing the control code to be adjusted directly (or closer) to the locking condition in a smaller number of steps~\cite{Yang2007,Hossain2014,Jung2015,Yao2015,Hsieh2016,Kim2021}. An AD-DLL with jitter filtering, achieved through an injection-locked oscillator, is presented in~\cite{Hossain2014}. A locking time of three cycles is demonstrated using a two-step TDC. An AD-DLL with phase-tracing delay unit and gated ring oscillators~\cite{Hsieh2016}, achieves a locking time of five cycles for a wide frequency range (6.7~MHz to 525~MHz). In~\cite{Yang2007}, a variable successive approximation register (SAR) locking algorithm using a TDC is proposed. Although the locking time is improved, it still requires 56 cycles to lock at 200~MHz. 
However, due to the nature of AD-DLL, previous schemes trade off resolution and jitter with portability, desensitization to process, voltage, and temperature (PVT), and faster locking time. 

In this work, a fast-locking, wide frequency range, and high-resolution MM-DLL architecture is presented. The proposed architecture, implemented in a 3-nm FinFET CMOS process, achieves a locking time of eleven cycles over a wide frequency range from 533~MHz to 4.26~GHz by means of a binary search (BS) locking scheme. 
This DLL reduces the locking time by 5-10$\times$ compared to traditional coarse-fine schemes, widely used in MM-DLLs. The BS is implemented by a finite-state machine (FSM) with bounded number of steps; and thus, the locking time becomes deterministic. A deterministic locking time reduces latency during transitions from idle to active modes and minimizes both startup and low-power mode exit times.

To address bias overshoot and potential clock failure events, we introduce a toggle detection circuit that can identify stalled clock conditions. The DLL loop controller samples the output of the toggle detector after each code change and applies corrective actions without needing to restart the entire initialization process. Leveraging the excellent time resolution in advanced CMOS process nodes, we achieve a sub-ps resolution and static phase error of 0.73~ps. The presented DLL consumes 5.4~mW from a 0.75-V power supply at 4.26~GHz. 

The manuscript is organized as follows. In Section~\ref{sec:MM-DLL}, an overview of MM-DLL and the coarse-fine locking scheme is presented. The proposed MM-DLL architecture, implementation details, the BS scheme, and the toggle detector are presented in Section~\ref{sec:BS-DLL}.  Simulation and experimental results are presented in Section~\ref{sec:Experimental}. A comparison between the proposed DLL and prior art is discussed in Section~\ref{sec:Comparison}. Finally, conclusions and prospects are discussed in Section~\ref{sec:Conclusion}.

\section{Mixed-Mode DLLs}
\label{sec:MM-DLL}
MM-DLLs offer a balanced trade-off between digital and analog DLLs, by providing both the flexibility and scalability of digital control, while retaining the precision of analog delay adjustment~\cite{Yang2007}. Thus, MM-DLLs achieve better tolerance to PVT variations, high resolution, low jitter, and low phase error. 

\subsection{MM-DLL Basic Architecture}
\begin{figure}[!t]
    \centering
    \includegraphics[width=\columnwidth,trim={0.6cm 0.6cm 0.6cm 0.6cm},clip]{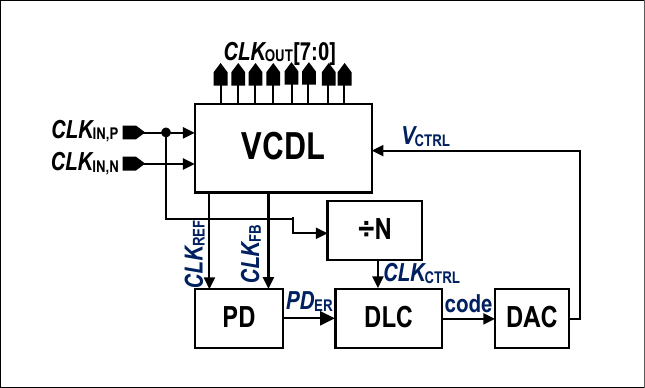}
    \caption{Block diagram of a generic MM-DLL.}
    \label{fig:DLL}
\end{figure}

A typical MM-DLL is composed by a differential VCDL, a PD, a digital loop control or filter (DLC), a clock divider ($\div N$), and a DAC as depicted in Fig.~\ref{fig:DLL}. The delay is dynamically controlled by the VCDL to guarantee that the $\Delta \phi$ is locked regardless of PVT and frequency variations~\cite{Moon2000}. The VCDL generates multiple equally-spaced phases of the input differential clocks (\textit{CLK}\textsubscript{IN,P}, \textit{CLK}\textsubscript{IN,N}). VCDLs are implemented as a chain of matched DEs with varying load capacitance and/or varying driving strength, and generates multiple equally-spaced clock phases (eight in this example, \textit{CLK}\textsubscript{OUT}[7:0]). The driving strength is modulated either by changing the power supply or by using a current-starved topology with bias voltage~\cite{Razavi2003}. The DAC generates the bias or supply voltage for the DE (\textit{V}\textsubscript{CTRL}). 

The phase difference $\Delta \phi$ is measured by the PD, which generates a binary output (\textit{PD}\textsubscript{ER}). As an example, \textit{PD}\textsubscript{ER} will be logic `1' if \textit{CLK}\textsubscript{REF} leads and logic `0' otherwise. The PD controls the DLC, which adjusts the DAC code to ensure that $\Delta \phi\approx 0$. Due to the binary nature of  \textit{PD}\textsubscript{ER} and the presence of jitter,  the DLL will dither around by at least $\pm 1$ code even after locking. The DLC is controlled by \textit{CLK}\textsubscript{CTRL}, a subdivided version of the input clock (\textit{CLK}\textsubscript{IN}). Running the DLC at a lower frequency is necessary as synthesizing digital logic at very high frequencies is typically challenging~\cite{Hossain2014}, and due to the relatively large settling time of the DAC compared to the input frequency. Consequently, the update rate, or loop bandwidth is primarily determined by the DLC and the DAC.

\begin{figure*}[!t]
    \centering
    \includegraphics[width=1\textwidth,trim={0.6cm 0.6cm 0.6cm 0.6cm},clip]{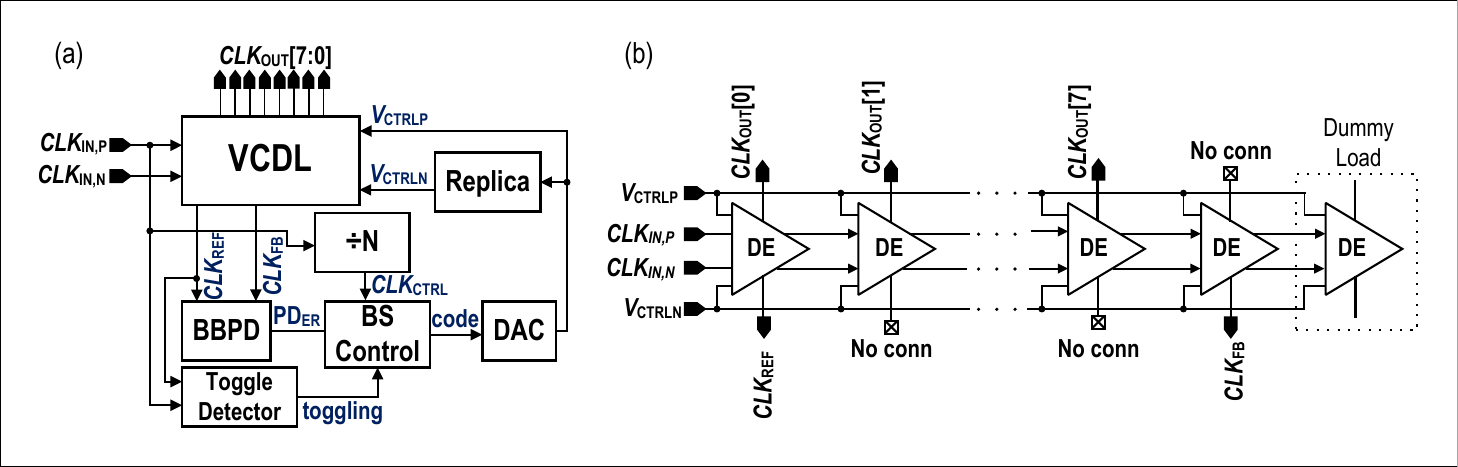}
    \caption{Proposed mixed-mode DLL architecture. (a) Block diagram of MM-DLL with binary search control in the digital loop and toggle detector. (b) Schematic diagram of the implemented voltage-controlled delay line (VCDL). }
    \label{fig:BS_DLL}
\end{figure*}

\subsection{Tradeoff Between Resolution vs. Locking Time }
The resolution in a MM-DLL ($\Delta T_{LSB}$) is determined by
\begin{equation}
    \Delta T_{LSB}=K_{VCDL}\cdot \Delta V_{LSB},
\end{equation}
where $K_{VCDL}$ is the gain of the VCDL ($K_{VCDL}=\partial \Delta \phi / \partial V_{ctrl}$), and $\Delta V_{LSB}$ is the least significant bit (LSB) resolution of the DAC. The DAC voltage resolution is $\Delta V_{LSB}=V_{FS}/2^N$, where $V_{FS}$ and $N$ are the full-scale voltage and number of bits, respectively. Then, the phase resolution ($\Delta \phi_{LSB}$) is expressed as
\begin{equation}
    \Delta \phi_{LSB}=2\pi f_{clkin} \Delta T_{LSB},
    \label{eq:phase_res}
\end{equation}
where $f_{clkin}$ is the frequency of \textit{CLK}\textsubscript{IN}. In real implementations, $K_{VCDL}$ is nonlinear, and varies with $V_{ctrl}$. Thus, the resolution of the DLL is not uniform throughout the whole DAC range. In other words, the VCDL presents non-zero differential non-linearity. 

For linear search, the locking time, $T_{Lock}$, is determined by 
\begin{equation}
T_{Lock}=\frac{\Delta \phi_{init}}{2\pi f_{clkin} \Delta T_{LSB}}\cdot \frac{1}{f_{clkctrl}},    
\label{eq:tlock_linear}
\end{equation}
where $\Delta \phi_{init}$ is the initial phase difference between the \textit{CLK}\textsubscript{FB} and the \textit{CLK}\textsubscript{REF}, and $f_{clkctrl}$ is the frequency of \textit{CLK}\textsubscript{CTRL}, respectively. Hence, as expected, $T_{Lock}$ is inversely proportional to $\Delta T_{LSB}$, $f_{clkin}$, and $f_{clkctrl}$, and proportional to $\Delta \phi_{init}$. In general, $f_{clkctrl}$ will be selected to be 4-10$\times$ lower than $f_{clkin}$ \cite{Razavi2020}.

\subsection{Coarse-Fine Locking}
The coarse-fine scheme uses a coarse step ($k$) during the initial locking procedure. When \textit{PD}\textsubscript{ER} changes polarity, the step size is reduced to a fine-grained step (\textit{e.g.}, 1). This scheme improves $T_{Lock}$ as the first change in the polarity of \textit{PD}\textsubscript{ER} will occur at approximately $\Delta \phi_{init}/(k\Delta \phi_{LSB} f_{clkctrl})$, where $k$ is the coarse step size. The worst-case scenario in terms of $T_{Lock}$ is that the overshoot leads to $\Delta \phi=k\Delta \phi_{LSB}$. Thus, if a step size of one LSB is used in the reminder of the locking process, $T_{Lock}$ will be 
\begin{equation}
    T_{Lock}=\frac{1}{f_{clkctrl}}\left( \frac{\Delta \phi_{init}}{k\Delta \phi_{LSB}} + k \right).
    \label{eq:tlock_coarse}
\end{equation}
For example, for $\Delta \phi_{init}\approx 2\pi$, $\Delta \phi_{LSB}=10~mRad$, $k=40$, and $f_{clkctrl}=1~GHz$, the locking time will be 46 cycles of \textit{CLK}\textsubscript{CTRL}. Note that the optimal value of $k$ would be $\sqrt{\Delta \phi_{init}/\Delta \phi_{LSB}}$ ($k=25$ for the previous example). However, since $\Delta \phi_{init}$ is unknown \textit{a priori} and $\Delta \phi_{LSB}$ is frequency dependent, $k$ must be chosen considering the input frequency range and PVT conditions. 

A major disadvantage of the coarse-fine scheme is the large overshoot in the bias voltage when large values of $k$ are used. An optimized version of this algorithm involves the use of a three-step coarse-fine scheme, starting with a large step size until the first polarity change in \textit{PD}\textsubscript{ER}, followed by a medium step size until the next polarity change, and finally using a step size of one to fine-tune the control until the locking condition is reached.

\section{Proposed Mixed-Mode Binary-Search DLL}
\label{sec:BS-DLL}

Motivated by the high time resolution requirements in state-of-the-art DDR and D2D physical layers, as well as the requirement of fast locking to lower latency, we propose a MM-DLL which uses a BS locking algorithm. In this section, we present the main architecture and enabling blocks. 

\subsection{Architecture}

\begin{figure*}[!t]
    \centering
    \includegraphics[width=1\textwidth,trim={0.6cm 0.6cm 0.6cm 0.6cm},clip]{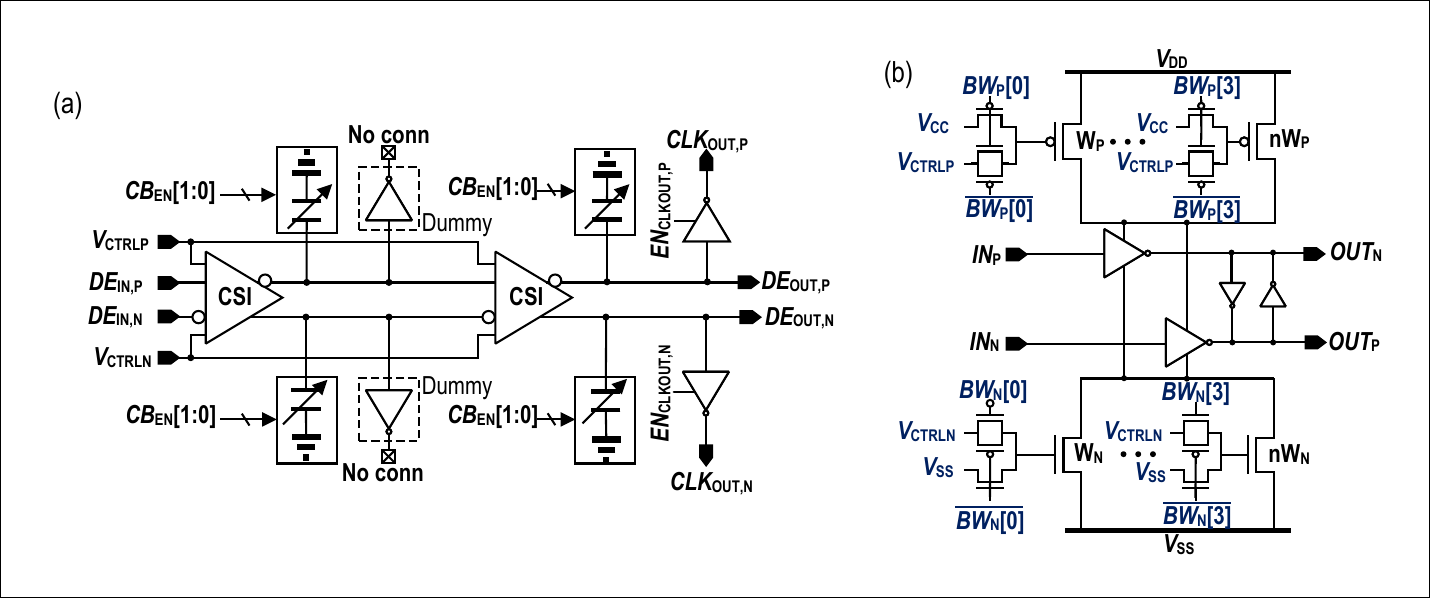}
    \caption{Schematic diagram of (a) the pseudo-differential delay element (DE) within the VCDL, and (b) the pseudo-differential current-starved inverter (CSI).}
    \label{fig:de_csi}
\end{figure*}

The proposed MM-DLL architecture is illustrated in Fig.~\ref{fig:BS_DLL}(a). It comprises a differential VCDL, bang-bang PD (BBPD), clock divider, DLC, and DAC. To improve locking time, the DLC employs a BS control mechanism, where an FSM generates DAC codes by navigating a binary tree. The BS control includes an accumulator, a shift register to adjust the step size, and registers to store the current and previous DAC codes. The FSM is clocked by a configurable clock divider, which supports divisions of $N=1,2,4,6,$ and $8$. This clock divider reduces the loop bandwidth, enhancing the timing margins of the synchronous circuits within the FSM, and minimizing output clock dithering. A toggle detector samples \textit{CLK}\textsubscript{REF} and raises a flag (\textit{toggling}) in the occurrence of a stalled clock condition.  

A 10-bit DAC produces the analog bias signal for the PMOS devices (\textit{V}\textsubscript{CTRLP}) in the VCDL, while a replica circuit generates the bias for the NMOS devices (\textit{V}\textsubscript{CTRLN}). These two bias voltages modulate the delay of the VCDL. The VCDL generates eight clock outputs (\textit{CLK}\textsubscript{OUT}[7:0]), evenly spaced by $T_{clkin}/8$, where $T_{clkin}$ is the period of the input clock. The VCDL is implemented as a chain of ten DEs, as shown in Fig.~\ref{fig:BS_DLL}(b). The ninth DE generates \textit{CLK}\textsubscript{FB}, while the tenth DE acts as a dummy load for delay matching. The number of DEs is kept to the possible minimum, as the output jitter in DLLs is directly proportional to the number of DEs in the VCDL~\cite{Gholami2016}.

\subsection{Delay Element Design}
The DE consists of two of cascaded pseudo-differential current-starved inverters (PS-CSIs), each loaded by a bank of capacitors, as illustrated in Fig.~\ref{fig:de_csi}(a). The total load capacitance ($C_B$) is adjusted by a bank of capacitors, controlled by \textit{CB}\textsubscript{EN}[1:0]. The output of the second PS-CSI stage drives a tri-state buffer, enabled by \textit{EN}\textsubscript{CLKOUT,P} and \textit{EN}\textsubscript{CLKOUT,N}, for the \textit{CLK}\textsubscript{OUT,P} and \textit{CLK}\textsubscript{OUT,N}, respectively. The DE outputs (\textit{DE}\textsubscript{OUT,P} and \textit{DE}\textsubscript{OUT,N}) drive the next DE stage. The first PS-CSI stage is loaded by dummy tri-state buffers for delay matching. Since both PS-CSIs are matched and evenly loaded, the total delay the DE, $t_{d,DE}$, is the sum of the delays of both CSIs, \textit{i.e.,} $t_{d,DE}=2t_{d,CSI}$, where $t_{d,CSI}$ is the delay of the CSI.

Each PS-CSI consists of two main inverters, each having two current-limiting transistors (\textit{i.e.}, tails), NMOS-based and PMOS-based, which starve the current of the main inverters, regulating their drive strength. The tails are controlled by bias voltages \textit{V}\textsubscript{CTRLP} and \textit{V}\textsubscript{CTRLN}, as shown in Fig.~\ref{fig:de_csi}(b). These biases are nearly symmetrical to ensure a balanced duty cycle. 
Thus, $t_{d,CSI}$ depends on the effective resistance ($R_{eff}$) and $C_B$, given by:
\begin{equation}
    t_{d,CSI}\propto R_{eff}C_B~,
\end{equation}
where $R_{eff}$ is modeled as:
\begin{equation}
    R_{eff}=\frac{L/W}{k(V_{DD}-V_T)}+\frac{L/W_{tail}}{k(V_{CTRLN}-V_T)}.
\end{equation}
Here, $L$, $W$, $k$, and $V_T$ represent the transistor length, width, transconductance, and threshold voltage, respectively, while $W_{tail}$ denotes the width of the tail transistors.

The number of active tail branches is tunable via  \textit{BW}\textsubscript{N}[3:0] and \textit{BW}\textsubscript{P}[3:0], which selectively connect each branch to $V_{DD}$ ($V_{SS}$) or to \textit{V}\textsubscript{CTRLP} (\textit{V}\textsubscript{CTRLN}) for the pull-up (down) paths [see Fig.~\ref{fig:de_csi}(b)]. A static branch (not shown) is always active to provide a minimum pull-up/down path for the main inverters. Thus, the values of \textit{BW}\textsubscript{N}[3:0] and \textit{BW}\textsubscript{P}[3:0] tune $W_{tail}$. Their values are dynamically adjusted according to the PVT and $f_{clkin}$ operating conditions.

Cross-coupled inverters are used to maintain the pseudo-differential clock paths aligned, minimizing skew within the VCDL. These cross-coupled inverters are chosen to be six times smaller than the main inverters. This flexible delay stage implementation, coupled with high DAC resolution, enables the DLL to cover a wide frequency range---from 533~MHz to 4.26~GHz---while achieving a $\Delta T_{LSB}$ of less than 1~ps.

\subsection{Binary Search Locking Procedure}

As described by~(\ref{eq:tlock_linear}), the locking time using a linear search increases with $\Delta T_{LSB}$, making this approach impractical for high-resolution DLLs. In particular, high resolution significantly impacts the low-frequency range, leading to hundreds of cycles before reaching the locking condition. Although the coarse-fine scheme reduces $T_{Lock}$ as expressed in (\ref{eq:tlock_coarse}), it still falls short for latency-critical applications. Therefore, in this work, we propose a BS controller to address the long startup and locking procedures. Rather than performing a linear or polynomial search, the binary search begins at code zero (\textit{i.e.}, the lowest delay) to prevent harmonic locking and then jumps directly to the midpoint of the code range. Based on the polarity of \textit{PD}\textsubscript{ER}, the search increases or decreases the code, halving the step size at each iteration. Fortunately, the division by two in the digital domain is relatively inexpensive, requiring only shifting right the binary code within a shift register. The BS continues this process until a step size of one is reached, achieving the desired locking condition.

\begin{figure}[!t]
    \centering
    \includegraphics[width=\columnwidth,trim={0.6cm 0.6cm 0.6cm 0.6cm},clip]{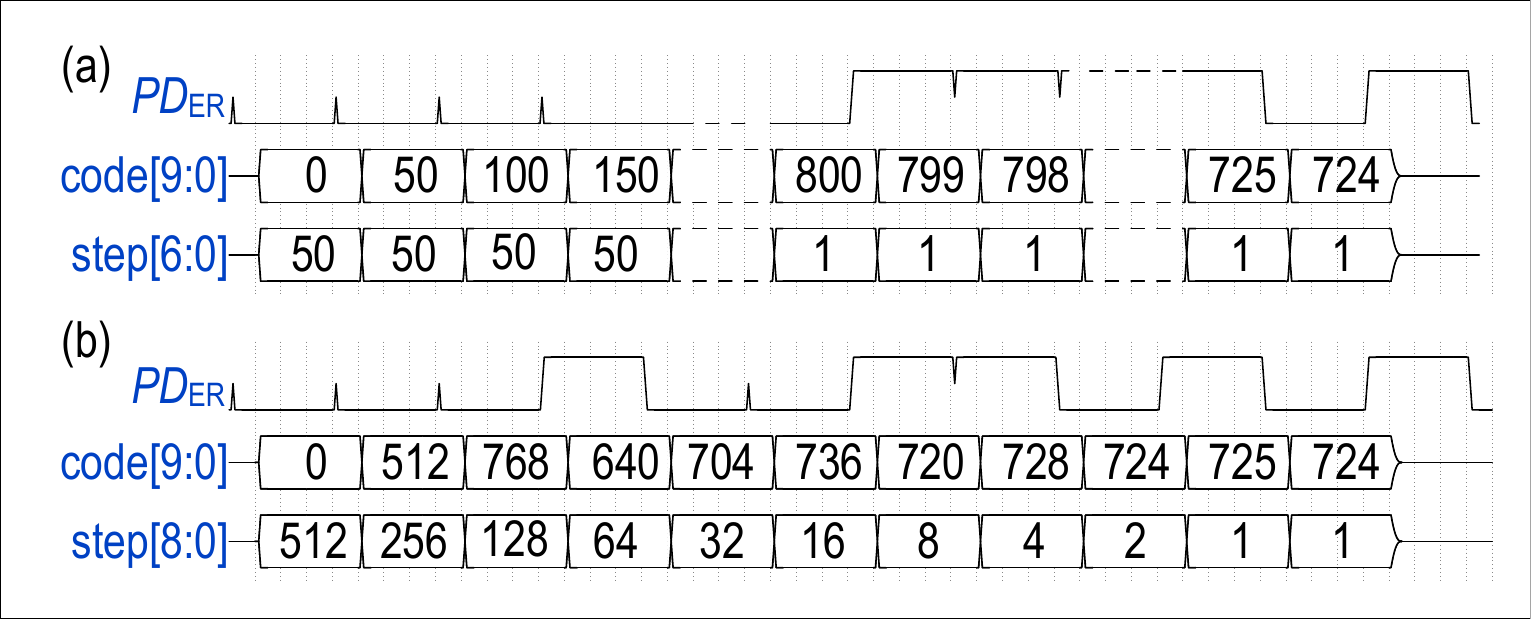}
    \caption{Locking algorithms example. (a) Coarse-fine scheme. (b) BS scheme.}
    \label{fig:BS_waveform}
\end{figure}

Examples of locking search procedures for the coarse-fine and the BS are depicted in Fig.~\ref{fig:BS_waveform}(a) and Fig.~\ref{fig:BS_waveform}(b), respectively. The coarse-fine algorithm begins with coarse step sizes of $k=50$ until a polarity change in \textit{PD}\textsubscript{ER}. Next, the search proceeds with step sizes of one until the locking condition is met. In contrast, the BS scheme navigates a binary tree. The step size starts at half the full range (\textit{i.e.}, $1024/2=512$). The step size is halved at each cycle of $f_{clkctrl}$. 
Once the step size reaches one, the DLL locks. In the absence of clock failure conditions, the locking process is completed in $\log_2{\left(2^B\right)}+1=B+1$ steps, where $B$ is the bit precision of the DAC ($B=10$ in this case). Therefore, the locking time becomes $T_{Lock}=(B+1)/f_{clkctrl}$, which is 5-10$\times$ faster than the coarse-fine scheme. 

\begin{figure}[!t]
    \centering
    \includegraphics[width=\columnwidth,trim={0.8cm 0.6cm 1.2cm 0.6cm},clip]{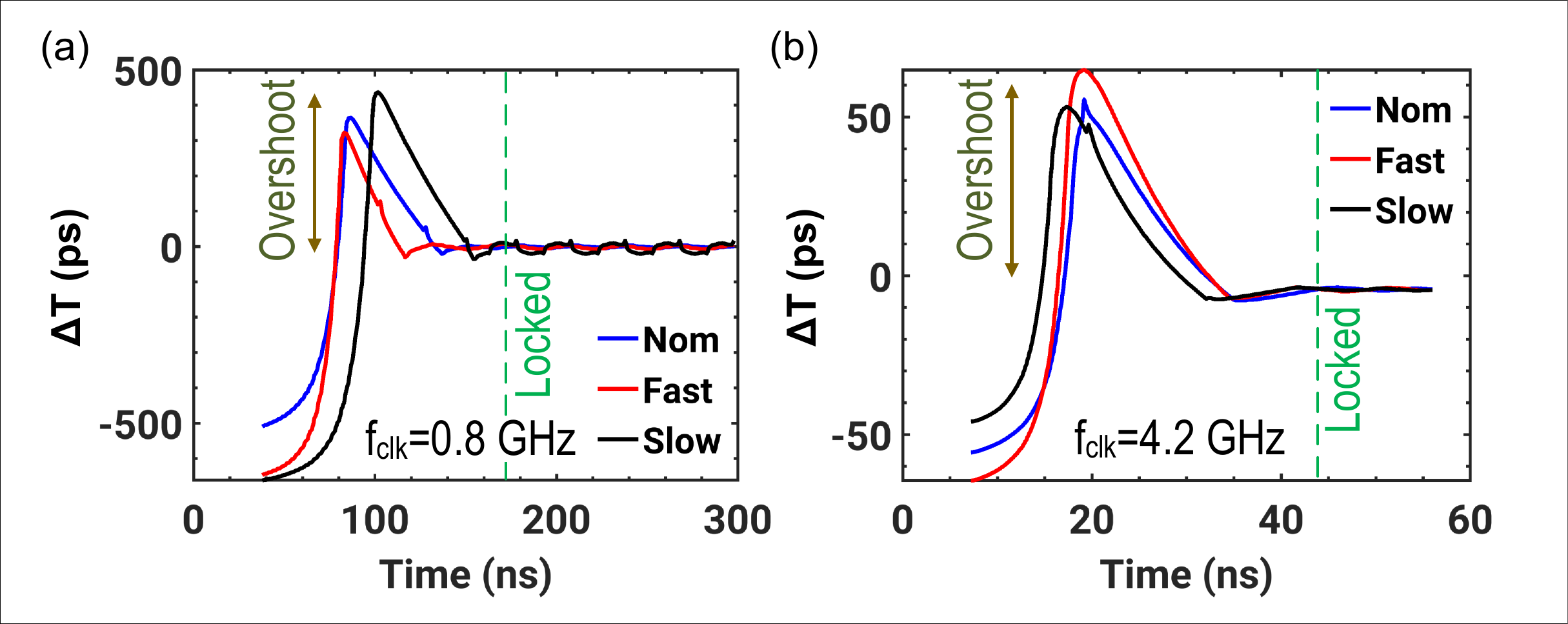}
    \caption{Simulation of the coarse-fine locking scheme for different PVT conditions. Change in $\Delta T$ for $f_{clkin}$ of (a) 800~MHz and (b) 4.26~GHz, at nominal (blue), fast (red), and slow (black) corners.}
    \label{fig:turbo_sim}
\end{figure}

For comparison, the proposed MM-DLL is tested with both the coarse-fine scheme and the BS algorithm. The simulation results of the startup locking process using the coarse-fine algorithm are shown in Fig.~\ref{fig:turbo_sim}(a) and Fig.~\ref{fig:turbo_sim}(b) for input frequencies of 800~MHz and 4.26~GHz, respectively, across three process corners. In particular, the skew between \textit{CLK}\textsubscript{REF} and \textit{CLK}\textsubscript{FB} ($\Delta T=2\pi f_{clkin}\Delta \phi$) initially rises sharply due to coarse steps, resulting in significant overshoot. When the polarity of \textit{PD}\textsubscript{ER} changes, finer steps are introduced until the DLL reaches lock, with $\Delta T$ dithering around zero in steady state. This dithering arises from the binary nature of the BBPD output, causing the DAC code to fluctuate by at least $\pm$1 code even in the locked state. Additionally, the $T_{Lock}$ is relatively long at both frequencies. Notably at 800~MHz, the locking process can extend beyond 140~ns, making it impractical for latency-critic applications.

\begin{figure}[!t]
    \centering
    \includegraphics[width=\columnwidth,trim={0.8cm 0.6cm 1cm 0.6cm},clip]{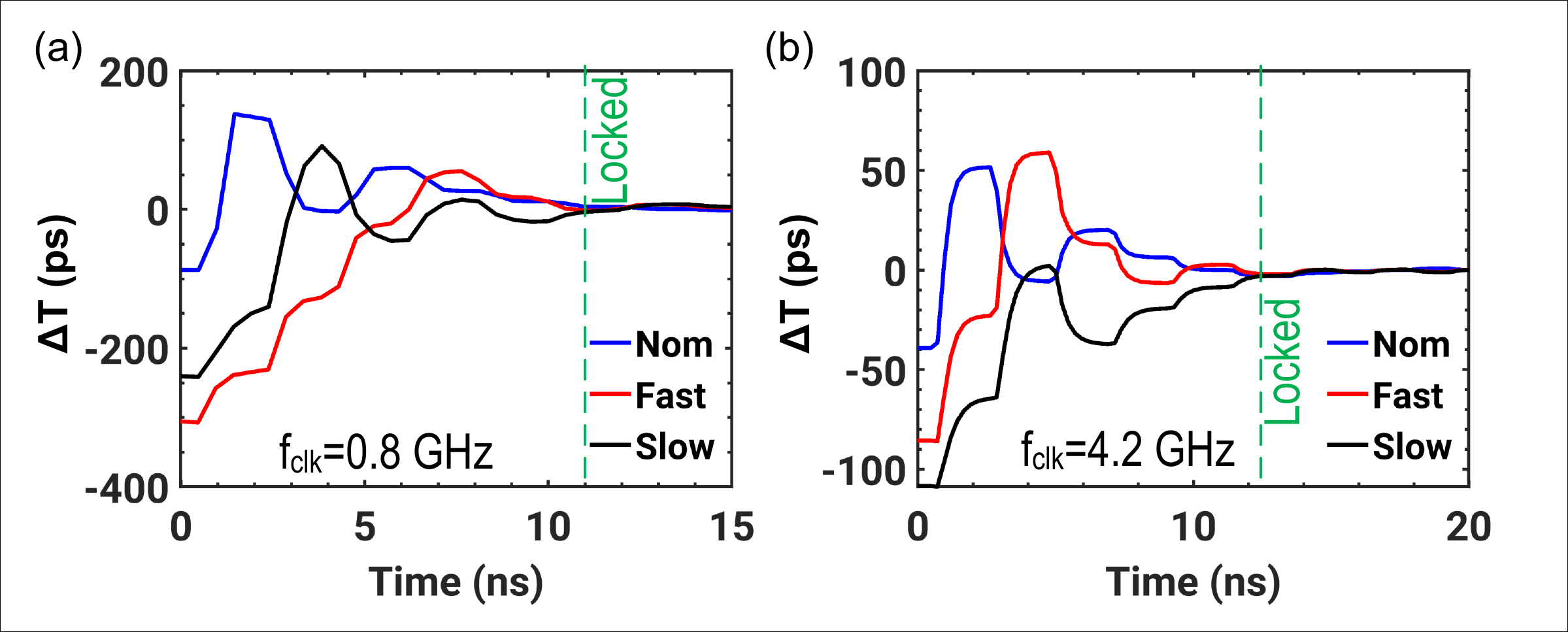}
    \caption{Simulation results of the BS locking scheme for different PVT conditions. Change of $\Delta T$ for $f_{clkin}$ of (a) 800~MHz and (b) 4.26~GHz, at nominal (blue), fast (red), and slow (black) operating conditions.}
    \label{fig:bs_sim}
\end{figure}

Simulations of the BS scheme for three corners are depicted in Fig.~\ref{fig:bs_sim}(a) and Fig.~\ref{fig:bs_sim}(b) for 800~MHz and 4.26~GHz, respectively. It can be observed that the DLL achieves lock within approximately 12.5~ns and 10.5~ns, for 800~MHz and 4.26~GHz, respectively, which corresponds to eleven cycles of \textit{CLK}\textsubscript{CTRL} (the simulation includes an initial reset and enable). This represents a reduction of 5-10$\times$ in the locking time compared to the coarse-fine scheme. Notably, the startup time remains consistent across PVT conditions, as the BS ensures deterministic locking time in the absence of clock failure events. The deterministic locking time is a key advantage of the proposed architecture. Even in worst-case scenarios with stalled clock conditions, the maximum locking time remains predictable. Additionally, the reduced locking time significantly enhances the cold boot and low-power mode exit latency in microprocessors, as well as in high-performance parallel interfaces, such as D2D, where latency is a critical design consideration.

\subsection{Clock Failure Detector}
\begin{figure}[!t]
    \centering
    \includegraphics[width=\columnwidth,trim={0.6cm 0.6cm 0.6cm 0.6cm},clip]{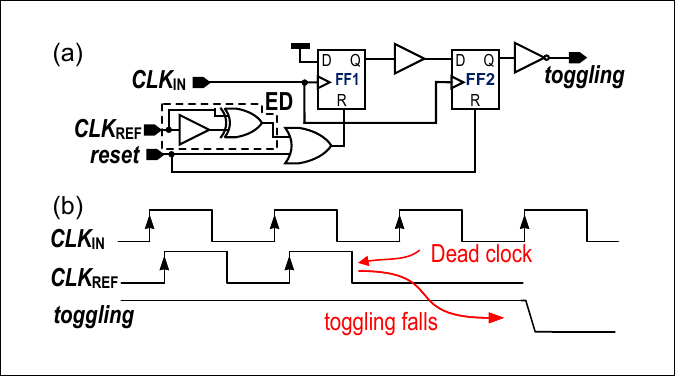}
    \caption{Toggle detector. (a) Schematic diagram. (b) Timing diagram.}
    \label{fig:toggle_detector}
\end{figure}

Although the BS algorithm significantly reduces $T_{Lock}$, it can still cause large bias overshoots due to the large step sizes at the beginning of the process, potentially leading to clock failure conditions. To address this, we implement a toggle detector circuit, shown in Fig.~\ref{fig:toggle_detector}(a). This custom digital circuit consists of two flip-flops (FFs), an edge detector (ED), buffers, and an inverter. 
The toggle detector operates as follows: Initially, both FFs are reset. As the input clock (\textit{CLK}\textsubscript{IN}) begins toggling, FF1 samples a constant logic `1' at the rising edge of \textit{CLK}\textsubscript{IN}. At any rising or falling edge of \textit{CLK}\textsubscript{REF}, the ED generates a pulse that clears FF1. Since the delay between \textit{CLK}\textsubscript{IN} and \textit{CLK}\textsubscript{REF} is small, FF1 will be cleared before FF2 can sample logic `1'. This keeps the detector output (\textit{toggling}) at logic `1'. However, if a stalled clock condition occurs and \textit{CLK}\textsubscript{REF} stops toggling, FF2 will sample the logic `1', switching the \textit{toggling} signal to logic `0'. This state persists until the FSM updates the code to a working value, at which point the toggle detector is reset along with the BBPD.

Waveforms illustrating the operation of the toggle detector are shown in Fig.~\ref{fig:toggle_detector}(b). The detector has a latency of less than one and a half \textit{CLK}\textsubscript{IN} cycles. Since \textit{CLK}\textsubscript{CTRL} operates at least at half the frequency of \textit{CLK}\textsubscript{IN}, this latency is effectively masked by the control clock cycle and does not affect the loop control bandwidth. The toggle detector can monitor different output clock pairs in the VCDL, with the condition that the phase difference between the clocks ($\Delta \varphi$) satisfies the condition 
\begin{equation}
2\pi f_{clkin}\Delta\varphi<T_{clkin}-t_{setup}-t_{R2Q}-t_{d_{buf}}-t_{d_{ED}}, 
\end{equation}
where $t_{setup}$ and $t_{R2Q}$ are the setup time and reset-to-out propagation delay of the FF, and $t_{d_{buf}}$ and $t_{d_{ED}}$ are the buffer and edge detector propagation delay, respectively. Consecutive clock phases can be compared and OR-ed to detect if any of the output clocks is not toggling.

\subsection{Binary Search FSM and Controller}
\begin{figure}[t!]
    \centering
    \includegraphics[width=\columnwidth,trim={0.6cm 0.6cm 0.6cm 1.5cm},clip]{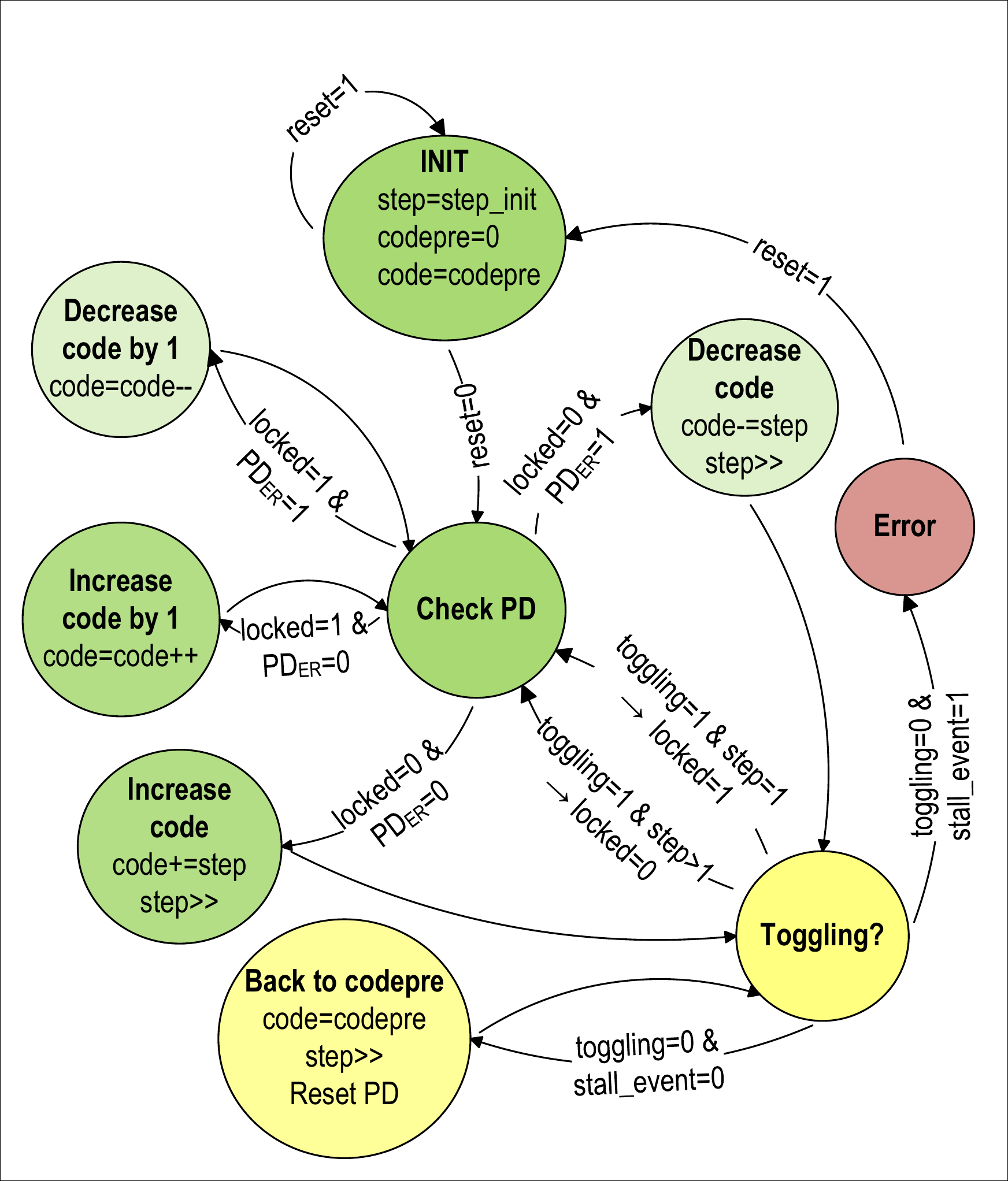}
    \caption{Simplified finite-state machine of the controller for the BS locking scheme.}
    \label{fig:fsm}
\end{figure}

A simplified version of the FSM for the BS controller is illustrated in Fig.~\ref{fig:fsm}. During reset and initialization, the step size is set to the middle of the code range, and the initial code is zero, although both the initial step and code are configurable. The previous code (codepre) is always stored in case a failure clock condition occurs.
At each iteration, the code is increased or decreased according to the polarity of \textit{PD}\textsubscript{ER}, and the step size is halved by right-shifting ($>>$) the logic `1' in the step size register. 
Then, the \textit{toggling} signal is sampled. If $toggling=1$, the BS process continues until the step size of one is reached, at which point locking is declared by raising a flag.

In the event of a stalled clock condition (\textit{toggling=0}), the code reverts to its last known working value, while the step size is halved to make a finer bias adjustment, and a \textit{stall\_event} flag is raised. During the next cycle of \textit{CLK}\textsubscript{CTRL} the code is updated to the previous working value plus or minus the smaller step size. If the stalled clock condition persists even with the finer adjustment, the FSM flags an error and enters a wait state until the DLL is reset. At this point, a new startup sequence is initiated with modified settings for the capacitor bank and the number of enabled PMOS/NMOS tails in the VCDL. A debug feature is also incorporated, enabling the FSM to freeze its operation, allowing the DAC code and clock outputs to be sampled at any given point for troubleshooting.

\begin{figure}[t!]
    \centering
    \includegraphics[width=\columnwidth,trim={1cm 0.8cm 0.8cm 0.6cm},clip]{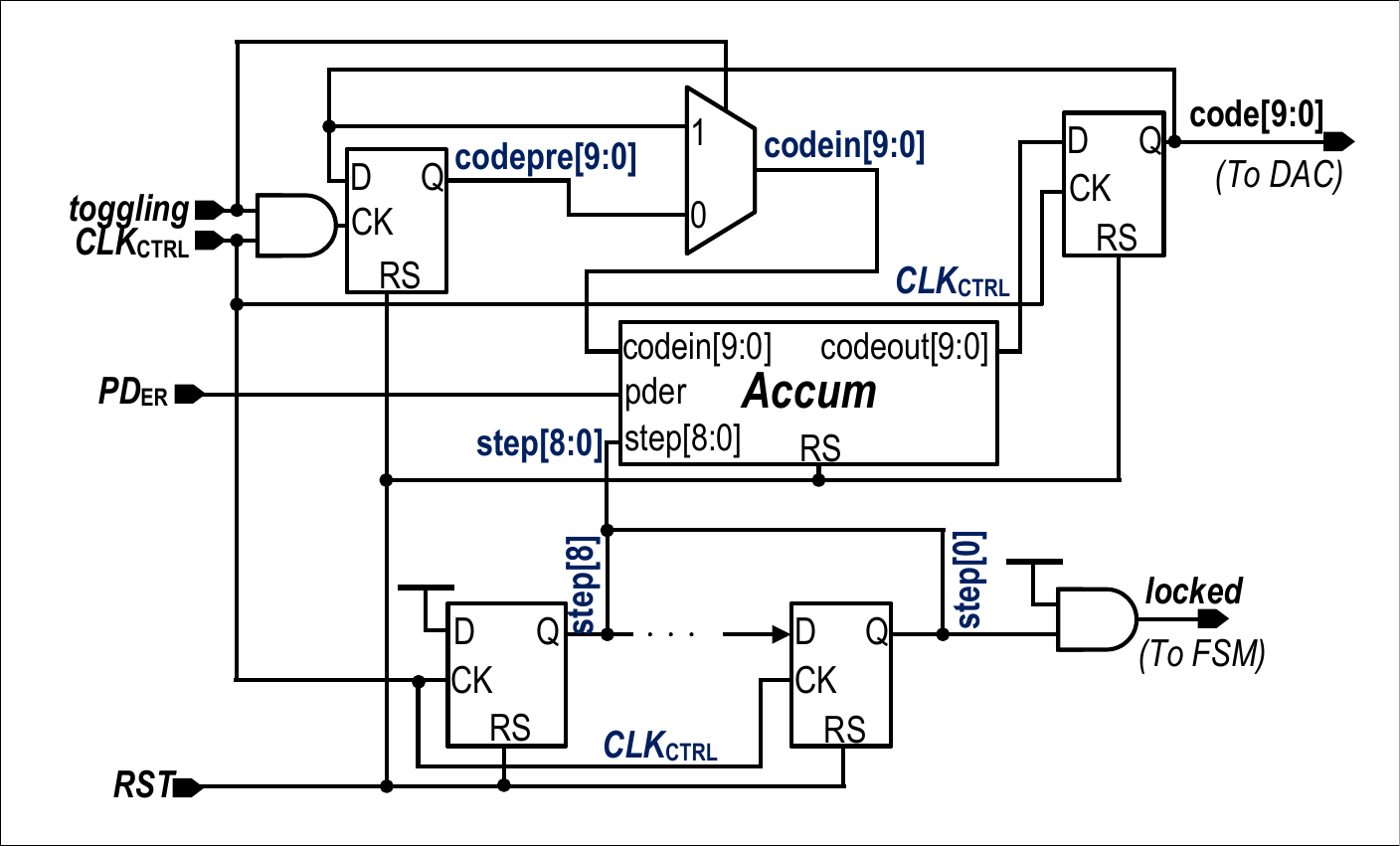}
    \caption{Simplified schematic of the BS controller.}
    \label{fig:bs_controller}
\end{figure}

A schematic representation of part of the BS controller is shown in Fig.~\ref{fig:bs_controller}. Two 10-bit flip-flop (FF) registers store the previous and current DAC code values. A shift register manages the BS step size adjustments by shifting logic `1' to the right. A 2:1 multiplexer (mux) selects either the current or previous code based on the \textit{toggling} status. The accumulator adjusts the 10-bit code by step[8:0] depending on the state of \textit{PD}\textsubscript{ER}. If \textit{toggling=0}, the FF holding the previous code is disabled, and the mux passes the last known working code to the accumulator. When step[0] is `1', the \textit{locked} flag is raised.

\section{Performance Evaluation and Experimental Results}
\label{sec:Experimental}
\begin{figure}[!t]
    \centering
    \includegraphics[width=0.8\columnwidth,trim={0.6cm 0.6cm 0.6cm 0.6cm},clip]{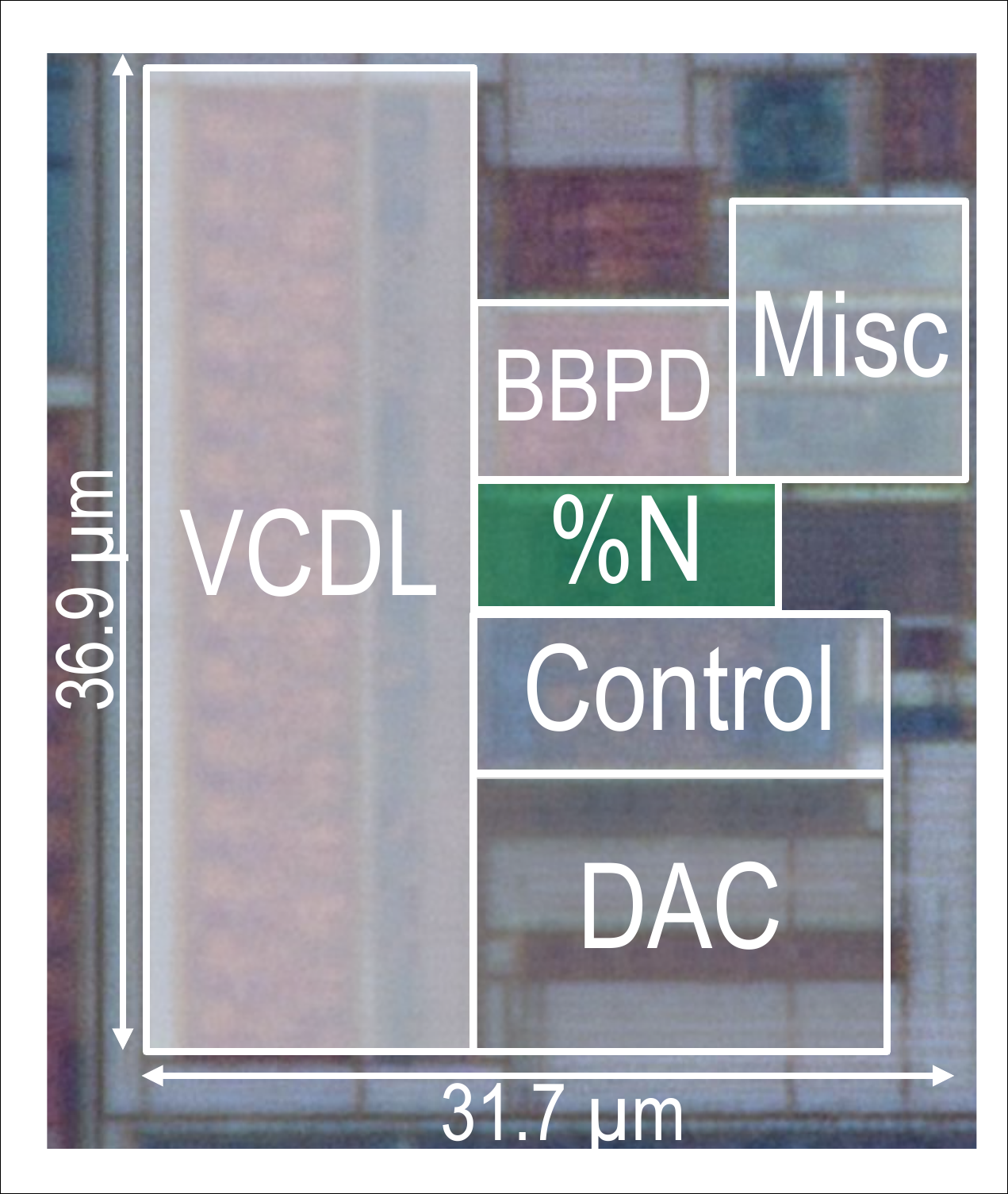}
    \caption{Micrograph of the proposed MM-DLL in a 3-nm FinFET process. The total area is $31.7\times 36.9 \mu m^2$.}
    \label{fig:micrograph}
\end{figure}

\begin{figure}[!t]
    \centering
    \includegraphics[width=0.9\columnwidth,trim={0.6cm 0.4cm 0.6cm 0.4cm},clip]{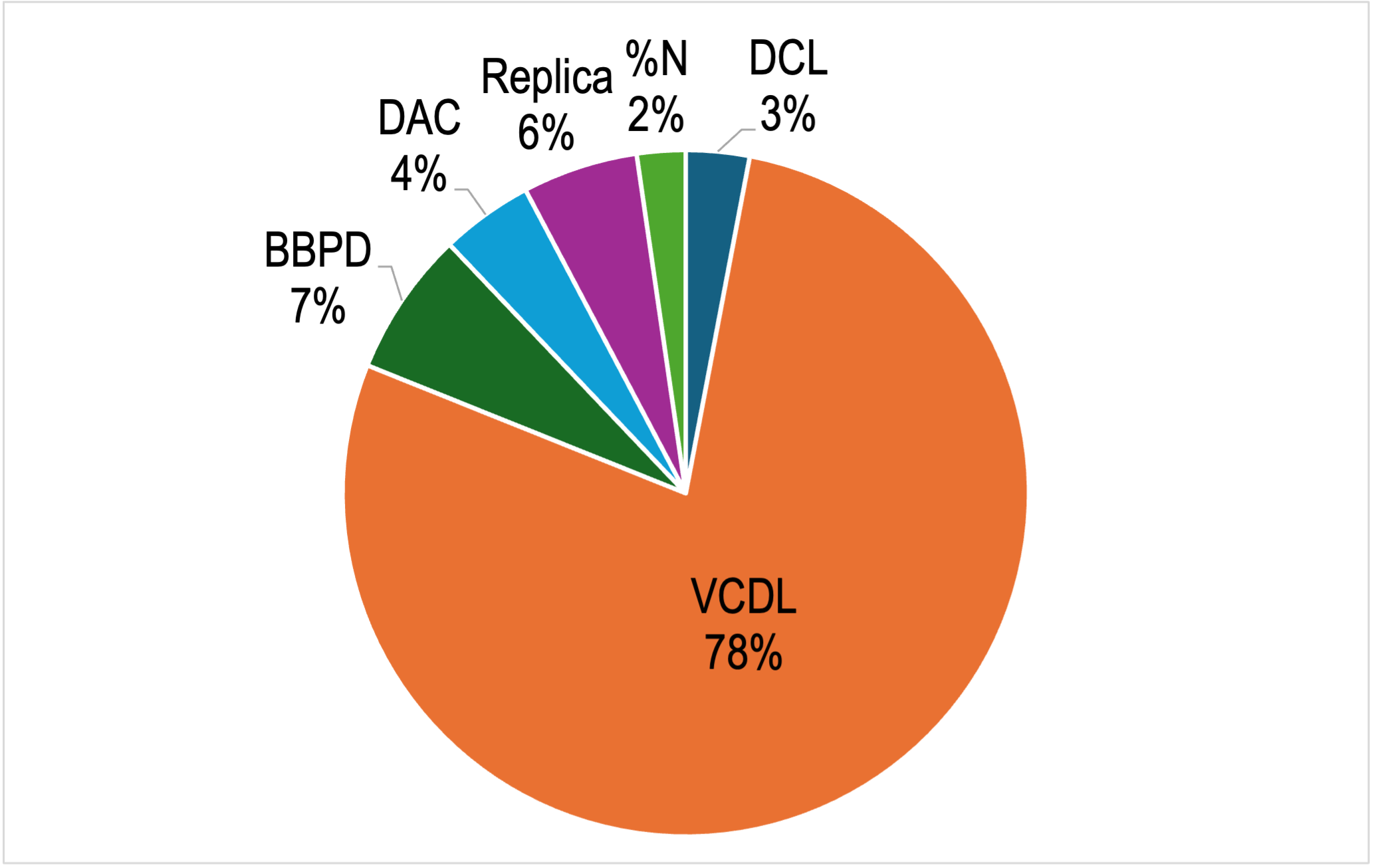}
    \caption{Power breakdown of the proposed MM-DLL.}
    \label{fig:power}
\end{figure}

The MM-DLL is implemented in an advanced 3-nm FinFET CMOS process, with a core area of $31.7\times36.9~\mu m^2$ as illustrated in the optical micrograph in Fig.~\ref{fig:micrograph}. This area includes the phase drivers that serve a whole data partition, which occupy an area of $10\times 36.9 \mu m^2$.  
The MM-DLL can operate across a nominal frequency range of 533~MHz to 4.26~GHz, constrained by the internal phase-locked loop (PLL) generating \textit{CLK}\textsubscript{IN,P} and \textit{CLK}\textsubscript{IN,N} rather than the DLL itself.
Achieving such a low frequency in a 3-nm process requires significant capacitive loading and/or aggressive dynamic voltage scaling to increase the delay of each DE stage. 
The MM-DLL consumes 5.4~mW from a 0.75-V power supply at $f_{clkin}=4.26~GHz$. A power breakdown is presented in Fig.~\ref{fig:power}. As expected, the VCDL accounts for almost 80\% of the power consumption. From the 5.4~mW, almost 1~mW is consumed by the phase drivers within the VCDL.

\begin{figure}[t!]
    \centering
    \includegraphics[width=0.9\columnwidth,trim={0.8cm 0.6cm 0.6cm 0.6cm},clip]{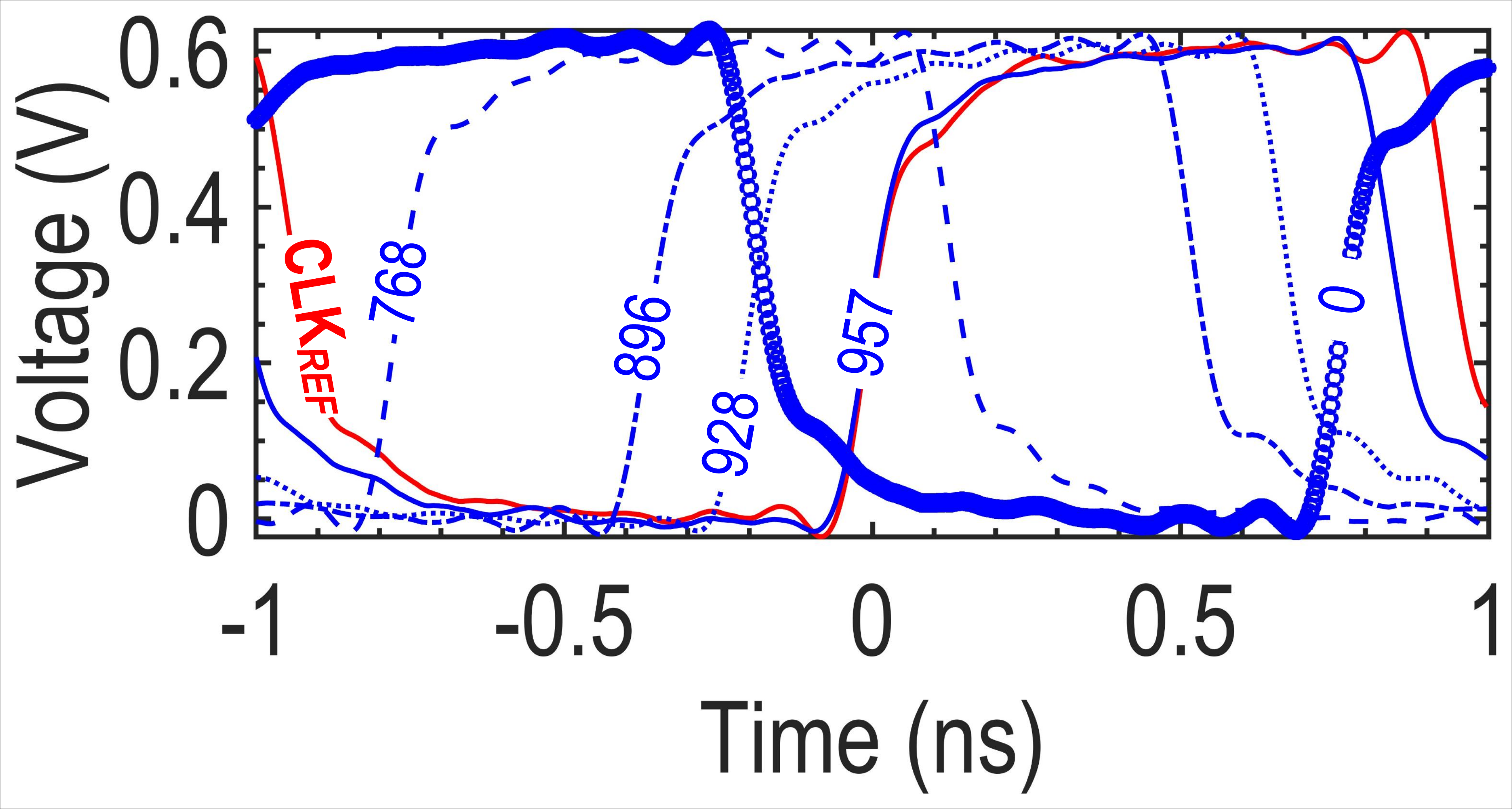}
    \caption{Experimental measurements of the BS locking process at 533~MHz. Transient waveforms of \textit{CLK}\textsubscript{REF} (red) vs. \textit{CLK}\textsubscript{FB} (blue, with different dash patterns representing each code). Locking is achieved at code 957.}
    \label{fig:bs_experimental_533MHz}
\end{figure}

\begin{figure}[t!]
    \centering
    \includegraphics[width=0.95\columnwidth,trim={0.8cm 0.6cm 0.6cm 0.6cm},clip]{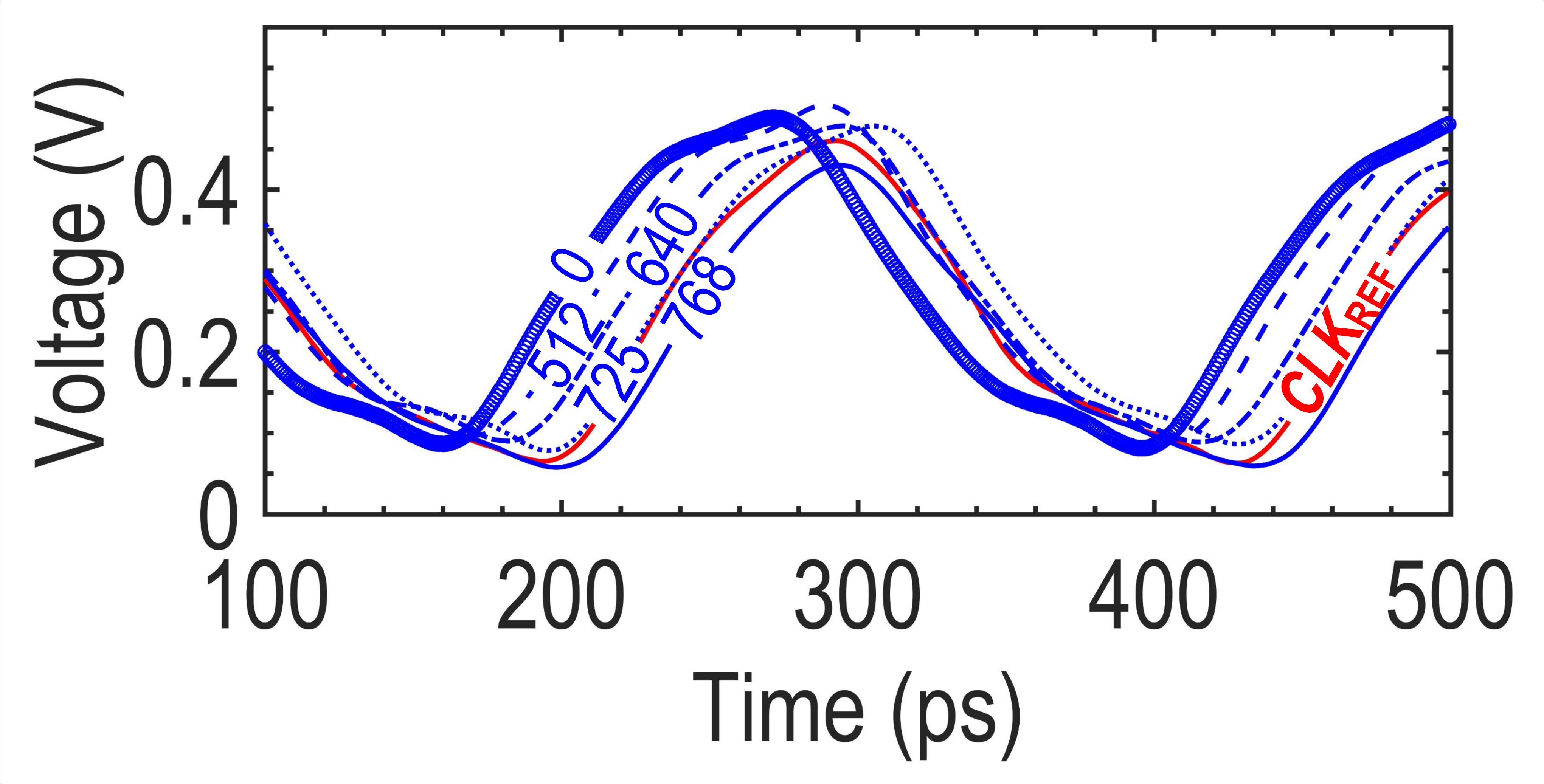}
    \caption{Experimental measurements of the BS locking process at 4.26~GHz. 
    Transient waveforms of \textit{CLK}\textsubscript{REF} (red) vs. \textit{CLK}\textsubscript{FB} (blue, with different dash patterns representing each code). Locking is achieved at code 725.}
    \label{fig:bs_experimental}
\end{figure}

\begin{figure}
    \centering
    \includegraphics[width=0.95\columnwidth,trim={0.8cm 0.6cm 0.6cm 0.6cm},clip]{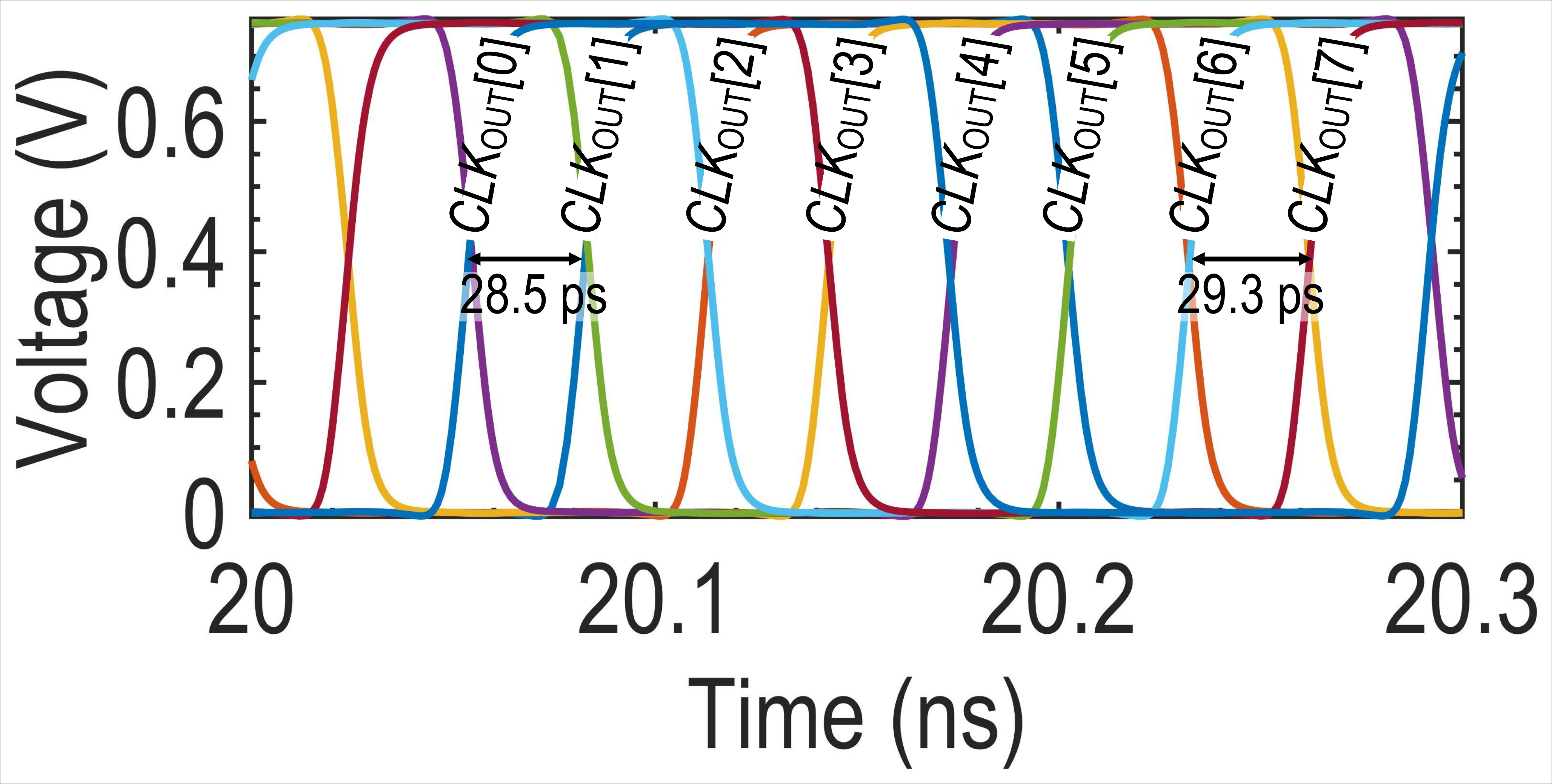}
    \caption{Simulation results of the eight output clock phases (\textit{CLK}\textsubscript{OUT}[7:0]) for $f_{clkin}=4.26~GHz$. The maximum phase error betweem output clocks is 0.8~ps.}
    \label{fig:phases}
\end{figure}

\begin{figure}
    \centering
    \includegraphics[width=\columnwidth,trim={0.8cm 0.6cm 0.6cm 0.6cm},clip]{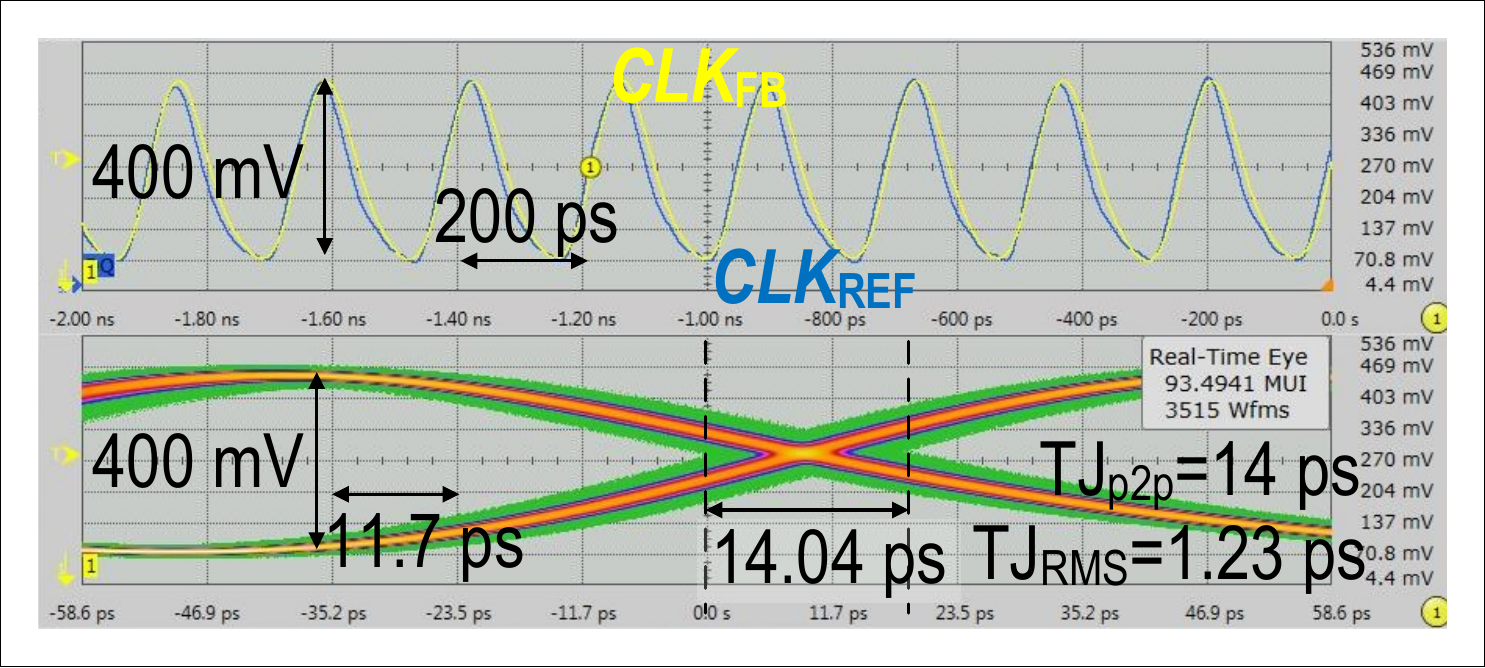}
    \caption{Experimental measurement of \textit{CLK}\textsubscript{FB} (yellow) and \textit{CLK}\textsubscript{REF} (blue), and eye diagram of \textit{CLK}\textsubscript{FB}. Measured $TJ_{p2p}$ and $TJ_{RMS}$ are 14~ps and 1.23~ps, respectively.}
    \label{fig:jitter}
\end{figure}

\begin{figure}
    \centering
    \includegraphics[width=\columnwidth,trim={0.6cm 0.6cm 0.8cm 0.6cm},clip]{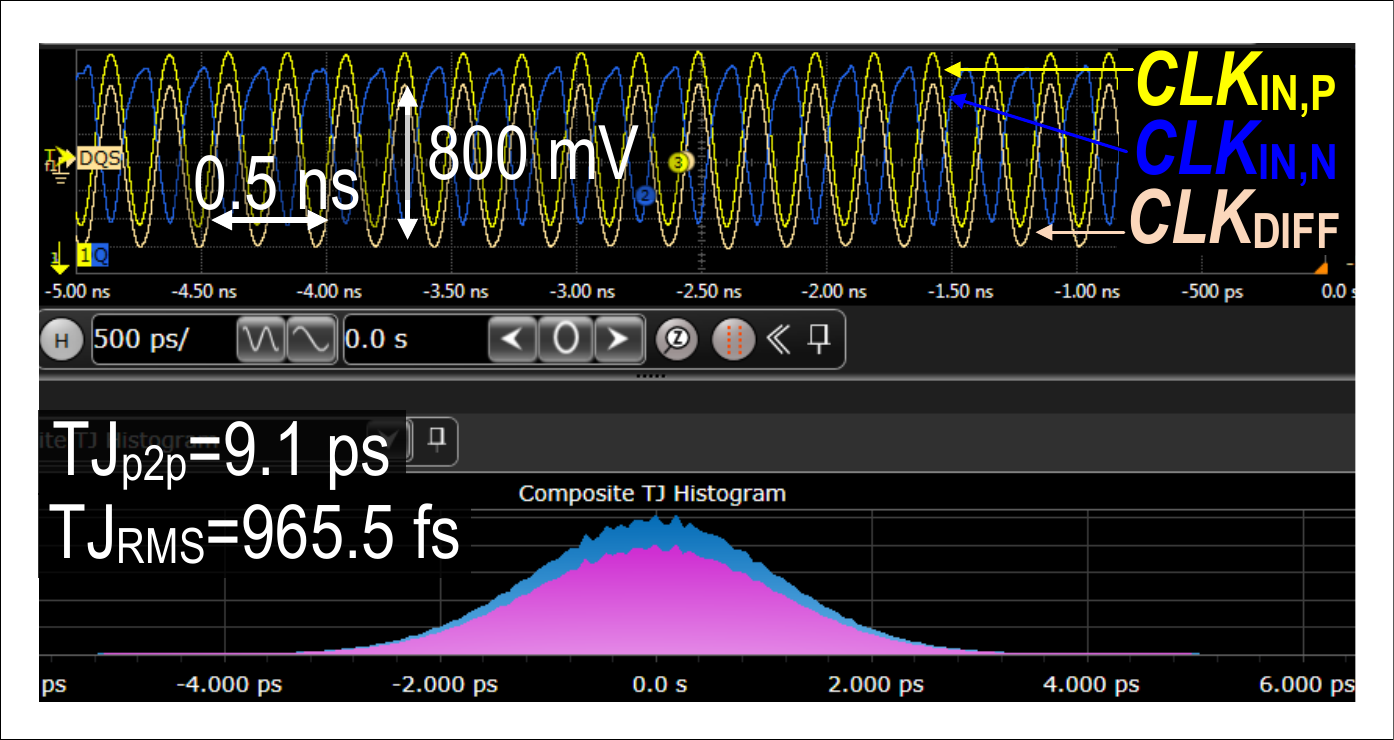}
    \caption{Experimental measurements of input differential clocks, \textit{CLK}\textsubscript{IN,P} (yellow) and \textit{CLK}\textsubscript{IN,N} (blue), and the differential signal (gray). The measured $TJ_{p2p}$ and $TJ_{RMS}$ are 9.1~ps and 965.5~fs, respectively.}
    \label{fig:jitter_diff}
\end{figure}

Experimental measurements of the BS locking algorithm for $f_{clkin}=533~MHz$ and $f_{clkin}=4.26~GHz$ are shown in Fig.~\ref{fig:bs_experimental_533MHz} and Fig.~\ref{fig:bs_experimental}, respectively. The clock signals are sampled from a digital debug chain that routes the DLL clocks to two I/O buffers, and the signals are measured on a development board using a high-speed oscilloscope configured with a 50-$\Omega$ input impedance. The phase shift between \textit{CLK}\textsubscript{REF} (red) and \textit{CLK}\textsubscript{FB} (blue) progresses in a BS pattern according to the DAC code. For clarity, only five steps of the locking sequence are shown. The measured phase resolution is 0.73 ps at 4.26~GHz, which represents the minimum achievable static phase error. The measured eye diagram of \textit{CLK}\textsubscript{FB} is shown in Fig.~\ref{fig:jitter}. The measured peak-to-peak jitter ($TJ_{p2p}$) and the RMS jitter ($TJ_{RMS}$) are 14~ps and 1.2~ps, respectively. These measurements include the jitter from the digital view path (used for debugging), with no de-embedding applied. The differential input clocks (Fig.~\ref{fig:jitter_diff}) show a $TJ_{p2p}$ of 9.1~ps, indicating that the proposed MM-DLL contributes only 4.9~ps of $TJ_{p2p}$. The simulated phase error between the output clock phases at 4.26~GHz is 0.8~ps, as shown in Fig.~\ref{fig:phases}. 

\section{Comparison}
\label{sec:Comparison}

\begin{table*}[!t]
    \centering
    \renewcommand{\arraystretch}{1.2}
    \captionsetup{singlelinecheck=true}
    \caption{\textsc{{Performance Comparison with Prior DLL Architectures}}}
    \label{tab:comparison}
    
   {\begin{tabular}{lC{1.5cm}C{1.5cm}C{1.5cm}C{1.5cm}C{1.5cm}C{1.5cm}C{1.5cm}} 
         \textbf{Reference} &  TCASI 20'~\cite{Angeli2020} & JSSC 21'~\cite{Park2021}  &  TCAS 15'~\cite{Jung2015} &  TVLSI 15'~\cite{Yao2015} & JSSC 16'~\cite{Hsieh2016} & TVLSI 21'~\cite{Tsai2021} & \textbf{This Work}\\ \hline \hline
         \textbf{Architecture}  & MM-DLL &  AD-DLL &  AD-DLL &  AD-DLL & AD-DLL & AD-DLL & MM-DLL \\ \hline
         \textbf{Locking scheme}  &  $\Sigma \Delta$ &  Coarse-fine &  Cyclic - locking loop &  TDC closed loop & TDC closed loop &  TDC closed loop & Binary - search \\ \hline
         \textbf{Technology (nm)}  &  65 &  28 &  65&  180 & 90 & 90 & \textbf{3} \\ \hline
         \textbf{Power Supply (V)} &  1.2 &  1.0 &  1.1&  1.8 & 1.2 & 1 & \textbf{0.75} \\ \hline
         \textbf{Operating Frequency (GHz)} &  2.8-4 &  1.3-4 &  0.4-0.8 &  0.06-1.2& 0.006--1.24 & 0.1--2.7& \textbf{0.533--4.26} \\ \hline
         \textbf{Frequency Ratio (FR)} & 0.25 & 1.02 & 0.67 & 1.81 & 1.98 & 1.86 & \textbf{1.55} \\ \hline
         \textbf{Locking time (ns)} &  1300 &  10 &  51.3 &  27.5 & 9.5 & 53 & \textbf{10.5} \\ \hline
         \textbf{Resolution (ps)} &  0.82 &  5.2 & 450 &  8.89 & 6 & 7.4 & \textbf{0.73} \\ \hline
         \textbf{TJ\textsubscript{RMS} / TJ\textsubscript{p2p} (ps)} & 0.84 / 2.28 &  1.86 / 12.5 &  4.8 / 26.1&  1.63 / 12.8 & 0.424 / 2.22 & 0.65 / 5 & \textbf{1.2 / 4.9} \\ \hline
         \textbf{Efficiency (mW/GHz)} &  0.74 &  1.28 &  4.4 &  13.2 & 11.29 & 18.3 & \textbf{1.27} \\\hline
         \textbf{Area (mm\textsuperscript{2})} & 0.0085 & 0.0044 & 0.017& 0.08& 0.032 & 0.089 & \textbf{0.0012}\\ \hline 
         \textbf{FOM\textsubscript{P} (pJ)} & 2.86	& 1.25 & 6.6 & 7.25 & 5.89 & 9.85 & \textbf{0.82} \\\hline
         \textbf{FOM\textsubscript{J} (pJ $\cdot$ ps)} & 1.6 & 15.9 & 114.8 & 168.9 & 25.1 & 91.5 & \textbf{6.2} \\\hline 
         \textbf{FOM\textsubscript{LR} (pJ $\cdot$ ns\textsuperscript{2})} & 0.76 & 0.066 & 101.5 & 3.2 & 0.65 & 7.2 & \textbf{0.010} \\\hline
         \hline
    \end{tabular}}
\end{table*}

A comparison between the proposed MM-DLL and previously reported DLL architectures (with an emphasis on fast-locking DLLs) is listed in Table~\ref{tab:comparison}. The proposed MM-DLL in 3-nm FinFET CMOS noticeably achieves a low locking time of 10.5~ns, without relying on TDC-based locking loops. Furthermore, it delivers a state-of-the-art time resolution of 0.73~ps, maintaining a high power efficiency of 1.27~mW/GHz, second only to~\cite{Angeli2020}. Notably, the phase drivers contribute with $\sim$1~mW to the total power, and the parasitic capacitance from the capacitor bank, necessary for low-frequency operation, slightly degrades power efficiency.
The proposed MM-DLL also demonstrates outstanding $TJ_{p2p}$ and $TJ_{RMS}$ performance compared to previous works. These figures are expected to improve further if the MM-DLL is characterized as a standalone component rather than within an SoC, as the current jitter measurements include contributions from the digital view path.

A frequency-range (FR) ratio is defined as in~\cite{Yang2021}:
\begin{equation}
    FR=\frac{2(f_{max}-f_{min})}{f_{max}+f_{min}},
    \label{eq:FR}
\end{equation}
where $f_{max}$ and $f_{min}$ represent the maximum and minimum operating frequencies, respectively. The proposed MM-DLL achieves a large FR of 1.55. Three figure-of-merits (FOMs) are defined for a fair comparison between the different works. Firstly, a power FOM is defined similar to~\cite{Tsai2021}:
\begin{equation}
    \label{eq:FOMp}
    FOM_P= \frac{P}{f_{max} \cdot FR}~[J],
\end{equation}
where $P$ is the power consumption. Secondly, a FOM which considers the locking time and resolution is defined as:
\begin{equation}
    \label{eq:FOMDLL}
    FOM_{LR}= FOM_P \cdot T_{Lock} \cdot \Delta T_{LSB} ~[J\cdot s^{2}].
\end{equation}
Finally, a FOM taking into account jitter is defined as:
\begin{equation}
    FOM_{J}=TJ_{p2p}\cdot \frac{P}{f_{max}} ~~[J \cdot s].
\end{equation}

\begin{figure}
    \centering
    \includegraphics[width=0.9\columnwidth,trim={1cm 1cm 1cm 0.8cm},clip]{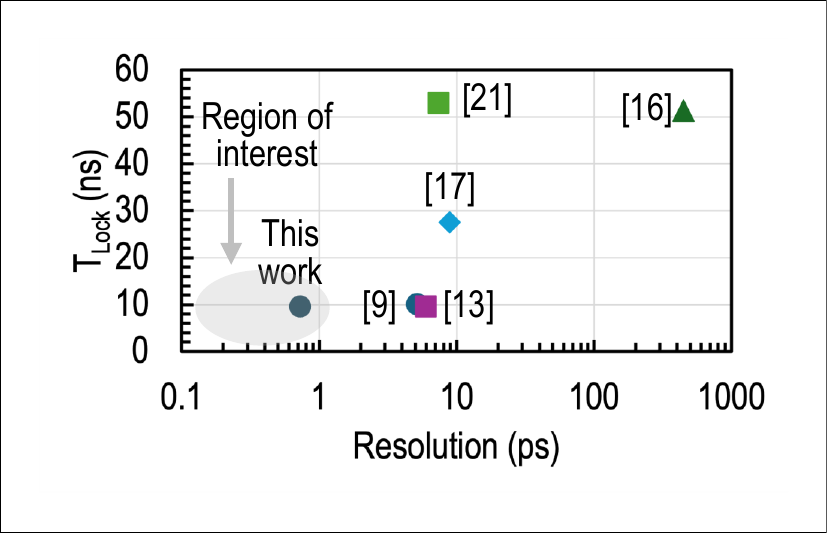}
    \caption{Locking time vs. resolution tradeoff for different DLL architectures. The proposed MM-DLL breaks the conventional trade-off between $T_{Lock}$ and resolution. }
    \label{fig:locking_res}
\end{figure}

The proposed MM-DLL achieves a FOM\textsubscript{P} of 0.82~pJ, representing at least a 5$\times$ improvement over prior art~\cite{Hsieh2016,Hossain2014,Park2021,Park2018, Angeli2019,Angeli2020,Tsai2021}. This is a major achievement, given that these prior works primarily rely on AD-DLL architectures. Furthermore, thanks to its high time resolution and BS locking algorithm, this design achieves state-of-the-art FOM\textsubscript{LR} of 0.01~pJ $\cdot$ ns\textsuperscript{2}, outperforming existing architectures by at least 6$\times$. The $T_{Lock}$ vs. resolution for various DLL architectures is shown in Fig.~\ref{fig:locking_res}. The proposed MM-DLL surpasses previous designs by achieving both high resolution and fast locking time, effectively breaking the conventional trade-off between these two key parameters. Furthermore, the achieved FOM\textsubscript{J} of 6.6~pJ$\cdot$ ps ranks among the best, second only to~\cite{Angeli2020}. 

\section{Conclusion}
\label{sec:Conclusion}

This paper presents a wide-range, fast-locking MM-DLL implemented in a 3-nm FinFET CMOS process. Although prior designs have often traded off timing resolution for faster locking, the proposed DLL leverages a BS controller to reduce locking time without compromising resolution or jitter. This architecture achieves a startup locking time of just eleven control clock cycles, improving overall system latency. To address potential clock failure conditions caused by bias overshoot, a toggle detector circuit is integrated into the control loop, enabling the BS FSM to recover the last functional DAC code and adjust the step size. The high-resolution DAC and BBPD further enable sub-ps time resolution and low jitter. With its combination of fast locking, high resolution, and robust clock failure recovery, the proposed MM-DLL is well suited for high-performance parallel I/O interfaces, such as DDR and D2D links, and other clock-sensitive applications in modern SoCs.

\section*{Acknowledgment}
The authors are grateful to Yael Kesselman for useful discussions on the FSM, to Eli Hamo for his support with the experimental setup and measurements, and to Igor Vornovitsky for the sample micrograph image.


%



\ifCLASSOPTIONcaptionsoff
  \newpage
\fi



%
\bibliographystyle{IEEEtran}

\bibliography{references}

%








\end{document}